\documentclass[aip,graphicx, amsmath,amssymb,reprint]{revtex4-1}
\usepackage{graphicx}
\usepackage{bm}
\usepackage{color}

\def\jnl@style{\it}
\def\aaref@jnl#1{{\jnl@style#1}}

\def\aaref@jnl#1{{\jnl@style#1}}

\def\aj{\aaref@jnl{AJ}}                   
\def\araa{\aaref@jnl{ARA\&A}}             
\def\apj{\aaref@jnl{ApJ}}                 
\def\apjl{\aaref@jnl{ApJ}}                
\def\apjs{\aaref@jnl{ApJS}}               
\def\ao{\aaref@jnl{Appl.~Opt.}}           
\def\apss{\aaref@jnl{Ap\&SS}}             
\def\aap{\aaref@jnl{A\&A}}                
\def\aapr{\aaref@jnl{A\&A~Rev.}}          
\def\aaps{\aaref@jnl{A\&AS}}              
\def\azh{\aaref@jnl{AZh}}                 
\def\baas{\aaref@jnl{BAAS}}               
\def\jrasc{\aaref@jnl{JRASC}}             
\def\memras{\aaref@jnl{MmRAS}}            
\def\mnras{\aaref@jnl{MNRAS}}             
\def\pra{\aaref@jnl{Phys.~Rev.~A}}        
\def\prb{\aaref@jnl{Phys.~Rev.~B}}        
\def\prc{\aaref@jnl{Phys.~Rev.~C}}        
\def\prd{\aaref@jnl{Phys.~Rev.~D}}        
\def\pre{\aaref@jnl{Phys.~Rev.~E}}        
\def\prl{\aaref@jnl{Phys.~Rev.~Lett.}}    
\def\pasp{\aaref@jnl{PASP}}               
\def\pasj{\aaref@jnl{PASJ}}               
\def\qjras{\aaref@jnl{QJRAS}}             
\def\skytel{\aaref@jnl{S\&T}}             
\def\solphys{\aaref@jnl{Sol.~Phys.}}      
\def\sovast{\aaref@jnl{Soviet~Ast.}}      
\def\ssr{\aaref@jnl{Space~Sci.~Rev.}}     
\def\zap{\aaref@jnl{ZAp}}                 
\def\nat{\aaref@jnl{Nature}}              
\def\iaucirc{\aaref@jnl{IAU~Circ.}}       
\def\aplett{\aaref@jnl{Astrophys.~Lett.}} 
\def\apspr{\aaref@jnl{Astrophys.~Space~Phys.~Res.}}
\def\bain{\aaref@jnl{Bull.~Astron.~Inst.~Netherlands}} 
\def\fcp{\aaref@jnl{Fund.~Cosmic~Phys.}}  
\def\gca{\aaref@jnl{Geochim.~Cosmochim.~Acta}}   
\def\grl{\aaref@jnl{Geophys.~Res.~Lett.}} 
\def\jcp{\aaref@jnl{J.~Chem.~Phys.}}      
\def\jgr{\aaref@jnl{J.~Geophys.~Res.}}    
\def\jqsrt{\aaref@jnl{J.~Quant.~Spec.~Radiat.~Transf.}}
\def\memsai{\aaref@jnl{Mem.~Soc.~Astron.~Italiana}}
\def\nphysa{\aaref@jnl{Nucl.~Phys.~A}}   
\def\physrep{\aaref@jnl{Phys.~Rep.}}   
\def\physscr{\aaref@jnl{Phys.~Scr}}   
\def\planss{\aaref@jnl{Planetary~and~Space~Sciences}}   
\def\procspie{\aaref@jnl{Proc.~SPIE}}   

\begin{document}




\preprint{}

\title{Comparison between hybrid and fully kinetic models of asymmetric magnetic reconnection: coplanar and guide field configurations} 








\author{Nicolas Aunai}
\email[]{nicolas.aunai@nasa.gov}
\affiliation{Space Weather Laboratory, Code 674, NASA Goddard Space Flight Center, Greenbelt, Maryland 20771, USA}

\author{Michael Hesse}
\affiliation{Space Weather Laboratory, Code 674, NASA Goddard Space Flight Center, Greenbelt, Maryland 20771, USA}

\author{Seiji Zenitani}
\affiliation{National Astronomical Observatory of Japan, 2-21-1 Osawa, Mitaka, Tokyo 181-8588, Japan}

\author{Maria Kuznetsova}
\affiliation{Space Weather Laboratory, Code 674, NASA Goddard Space Flight Center, Greenbelt, Maryland 20771, USA}

\author{Carrie Black}
\affiliation{Space Weather Laboratory, Code 674, NASA Goddard Space Flight Center, Greenbelt, Maryland 20771, USA}

\author{Rebekah Evans}
\affiliation{Space Weather Laboratory, Code 674, NASA Goddard Space Flight Center, Greenbelt, Maryland 20771, USA}

\author{Roch Smets}
\affiliation{Laboratoire de Physique des Plasmas, Universite Pierre et Marie Curie, Ecole polytechnique, route de Palaiseau, 91128 Palaiseau Cedex}





\date{\today}

\begin{abstract}
Magnetic reconnection occurring in collisionless environments is a multi-scale process involving both ion and electron kinetic processes. Because of their small mass, the electron scales are difficult to resolve in numerical and satellite data, it is therefore critical to know whether the overall evolution of the reconnection process is influenced by the kinetic nature of the electrons, or is unchanged when assuming a simpler, fluid, electron model. This paper investigate this issue in the general context of an asymmetric current sheet, where both the magnetic field amplitude and the density vary through the discontinuity. A comparison is made between fully kinetic and hybrid kinetic simulations of magnetic reconnection in coplanar and guide field systems. The models share the initial condition but differ in their electron modeling. It is found that the overall evolution of the system, including the reconnection rate, is very similar between both models. The best agreement is found in the guide field system, which confines particle better than the coplanar one, where the locality of the moments is violated by the electron bounce motion. It is also shown that, contrary to the common understanding, reconnection is much faster in the guide field system than in the coplanar one. Both models show this tendency, indicating that the phenomenon is driven by ion kinetic effects and not electron ones.
\end{abstract}

\pacs{}

\maketitle 

\section{Introduction}

Magnetic reconnection is a universal plasma phenomenon enabling large scale plasma structures frozen in the magnetic field to change their magnetic connectivity while transferring a substantial part of the magnetic energy stored in current sheets into the thermal and bulk kinetic energy of the surrounding plasma\citep{Priest:2000wm,Yamada:2010im}. In most systems where reconnection is thought to play a key role, the amount of magnetic flux reconnected per time unit, the so-called reconnection rate, appears as a critical parameter for their large scale evolution. Understanding which physical mechanism controls it, and at which scale, is therefore an important issue from the fundamental and modeling viewpoint. In collisionless systems, the vastly different inertia of electrons and ions makes magnetic reconnection a multiscale process\citep{Birn:2007vz,Yamada:2010im}. Because it creates accelerated and heated flows and macroscopically changes the transport of the plasma in magnetized systems, magnetic reconnection can be thought as a fluid phenomenon. However, the collisionless nature of the systems in which it occurs involves kinetic processes for which the importance regarding this large scale evolution is still poorly understood. Although the question really concerns both species, knowing to what extent the kinetic behavior of the electrons plays a role in controlling the overall dynamics is particularly critical to understand, observe/measure and model the reconnection process, since it concerns scales that are, from all those viewpoints, very difficult to resolve, in comparison to ion scales.

Over the years, several studies have addressed this issue and different conclusions have been reached. On one hand, electron kinetic physics is seen as the fundamental way to break the flux freezing constraint in collisionless systems, but is thought not to affect the global process, which is rather controlled by the coupling of ion dynamics and electron fluid physics\citep{2001JGR...106.3715B,2008PhPl...15d2306D}. On the other hand, electron kinetic physics is seen as a key ingredient, which can make the whole process unsteady by enabling the frequent production of magnetic flux ropes\citep{2006PhPl...13g2101D,2007GeoRL..3413104K,2011NatPh...7..539D}. The applicability of our understanding is, however, most of the time limited by the assumption of an initial symmetry across the current sheet, which might be adequate for the modeling of some environments like the Earth magnetotail, but seems oversimplified in most other current layer systems like the Earth magnetopause, the solar wind and solar coronal loops. The few numerical studies which focused on kinetic reconnection in asymmetric current sheet, have revealed a great number of new features challenging our current understanding of magnetic reconnection\citep{2000JGR...10523179N,2000JGR...10525171X,2003JGRA..108.1218S,Pritchett:2008ef,2008AnGeo..26.2471T,2009JGRA..11411210P,2009PhPl...16h0702P,2010JGRA..11510223M,2011SSRv..158..119M}. Among them, recent studies have revealed a rich variety of electron scale kinetic processes\citep{2003JGRA..108.1218S,Pritchett:2008ef,2009PhPl...16h0702P,2009JGRA..11411210P,2011SSRv..158..119M}. Surprisingly, many of them were not confined to the reconnection site as one could have expected, but rather expand over large regions of the reconnection exhaust. Their omnipresence raises the question of whether including electron kinetic physics is a mandatory requirement for an adequate modeling of large scale reconnection of magnetic field in these more general systems, and to what extent neglecting it changes the overall dynamics. Along the same lines, the initial condition of kinetic simulations of asymmetric reconnection models is never at a kinetic steady state but in a fluid equilibrium, which very rapidly evolves toward a self-consistent quasi-steady state with an internal structure very different from the prescribed one\citep{Pritchett:2008ef}. The role of kinetic electrons in the establishment of such structure is unknown and it is important to understand to what extent neglecting this physics impacts the structure of the current sheet where reconnection develops.

In this paper, we focus on the role of electron kinetic physics in asymmetric magnetic reconnection. To do so, we present the results of two kinetic simulations sharing the same initial configuration but differing in the modeling of the electron population. One code solves the complete Vlasov-Maxwell system, and is commonly referred as a fully kinetic Particle-In-Cell (PIC) code\citep{Birdsall:2004vm}, while the other one assumes the electrons behave as an isothermal fluid without bulk inertia and the ions are treated as particles, and is usually referred as a hybrid PIC code\citep{lipatov2002}. These latter assumptions are of course not generally satisfied in real systems, but their consequences regarding the reconnection process are not well understood. The second part of this paper is dedicated to the description of the simulation methods with their respective physical assumptions. The third part describes the initial conditions used in this work and investigates to what extent the kinetic nature of electrons impacts the initial internal structure of the current sheet after the re-configuration of the initial fluid equilibrium. The fourth part presents the reconnection rate obtained from both simulations. The fifth part discusses the structure of the current sheet and the sixth and last section summarizes and discuss our findings.

\section{Numerical models}
This section describes the numerical models used in this study. The fully kinetic model treats both electrons and ions as a collection of particles feeling the electric $\mathbf{E}$ and magnetic $\mathbf{B}$ fields. The force they feel is described by eq. (\ref{eq:lorentz}), where $m_s$, $q_s$ and $\mathbf{v_s}$ are the mass, charge and velocity of the particle of species $s$. The electromagnetic fields are obtained by solving the Maxwell-Ampere (\ref{eq:ampere}) and Maxwell-Faraday (\ref{eq:faraday}) equations, where $\mathbf{j}$ is the total electrical current density, $c$ is the speed of light in vacuum and $\mu_0$ the magnetic permeability of vacuum. The details of the algorithms have already been explained in a previous paper \citep{2001JGR...106.3721H}. For the runs presented in this paper, we have used an electron to ion mass ratio $m_e/m_i=1/25$ and the ratio between the plasma frequency and the electron cyclotron frequency is $\omega_{pe}/\omega_{ce}=4$.

\begin{eqnarray}
m_s\frac{d\mathbf{v}_s}{dt} & = & q_s\left(\mathbf{E} + \mathbf{v}_s\times\mathbf{B}\right)\label{eq:lorentz}\\
\frac{1}{c^2}\frac{\partial \mathbf{E}}{\partial t} & =  & \boldsymbol{\nabla}\times\mathbf{B} -\mu_0\mathbf{j}\label{eq:ampere} \\
\frac{\partial \mathbf{B}}{\partial t} & = & -\boldsymbol{\nabla}\times\mathbf{E}\label{eq:faraday}
\end{eqnarray}

In the hybrid code, only the ions are treated as a collection of particles. Electrons, on the other hand, behave as a fluid, i.e. no kinetic behavior is included in their modeling. Furthermore, because of their small mass compared to ions, we also neglect quasi-neutrality scales and assume the electron density equals the ion one. Consistently, the displacement current is neglected in Maxwell-Ampere equation (\ref{eq:ampere}). The electron bulk inertia is also neglected. Their fluid momentum equation is thus used to update the electric field, based on these approximations (\ref{eq:ohm}), where $\mathbf{v}_e$ is the electron fluid velocity, $P_e$ their scalar pressure and $\mathbf{R}$ a dissipation term.

\begin{equation}\label{eq:ohm}
\mathbf{E} = -\mathbf{v}_e\times\mathbf{B} - \boldsymbol{\nabla}P_e + \mathbf{R}
\end{equation}

 In this study, we choose an hyperresistivity to model the dissipation, $\mathbf{R} = -\nu\boldsymbol{\nabla}^2\mathbf{j}$, where $\nu=5\ 10^{-4}$. Such dissipation is analogous to an electron viscosity mechanism, which is coincidentally similar to what occurs at the reconnection site in symmetric fully kinetic models \citep{2011SSRv..tmp...10H}. Its implementation in the modeling of asymmetric magnetic reconnection has furthermore been recently shown to have important consequences regarding the overall evolution of the system\citep{aunai2012a}. Details of the algorithms used in the hybrid code can be found in previous papers \citep{2007AnGeo..25..271S,2011JGRA..11609232A}.\\

Both models are have 2.5 dimensions. The plane in the fully kinetic model is called $x-z$ whereas in the hybrid model it is called $x-y$. The domain is considered periodic in the $x$ direction and closed by perfect conducting walls in the other one. The data presented in this paper is in normalized units. The magnetic field is normalized to an arbitrary magnetic field $B_0$, the density to an arbitrary density $n_0$, the time to the ion cyclotron period based on the ion mass $m_i$ and charge $q_i$, and on the magnetic field $B_0$. The lengths are normalized to the ion inertial length $\delta_i$ based on the density $n_0$. The electric field is therefore normalized to $V_AB_0$, where $V_A = B_0/\sqrt{n_0m_i\mu_0}$ is the characteristic Alfven speed in the system.

\section{Initial condition and equilibrium}
In this section, we describe the way the system studied in this paper is initialized. To simplify the problem, we assume the current sheet to be a one dimensional tangential discontinuity where the plasma properties and electromagnetic fields vary between two asymptotic and uniform values. We also assume the system to be in steady state before reconnection occurs. Following these hypotheses, the density $n$, temperature $T$ and magnetic field $\mathbf{B}$ profile must satisfy the 1D pressure balance condition (\ref{eq:1dpressurebalance}), where $k_B$ is the Boltzmann constant.

\begin{equation}\label{eq:1dpressurebalance}
nk_BT + \frac{B^2}{2\mu_0}=cst
\end{equation}

We choose the following profiles for the density (\ref{eq:densityinit}), the magnetic field (\ref{eq:magneticinit}) where $\lambda=0.5$. The ion temperature $T_i$ is obtained from equation (\ref{eq:1dpressurebalance}) and $T = T_i + T_e$, where $T_e=0.2T_i$ is the electron temperature.

\begin{eqnarray}
n(y) & = & 1 - \frac{1}{3}\left(S\left(y\right) +S\left(y\right)^2\right)\label{eq:densityinit}\\
B_x(y) &  = & \frac{1}{2} + S\left(y\right)\label{eq:magneticinit}\\
S\left(y\right) & = & \tanh\left(\frac{y-y_0}{\lambda}\right)\label{eq:shapefactor}
\end{eqnarray}

 \begin{figure}
 \includegraphics[width=\linewidth]{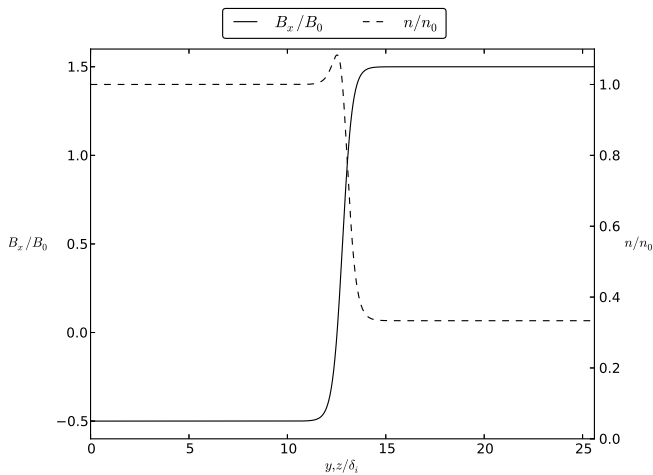}%
 \caption{\label{fig:init} The solid line represents the initial in-plane magnetic field profile given by eq. (\ref{eq:magneticinit}). The dashed line represents the initial density given by eq. (\ref{eq:densityinit}).}%
 \end{figure}

This initial condition has been used in previous studies \citep{2009JGRA..11411210P,Pritchett:2008ef,2011PhRvL.106s5003Z} and is illustrated on Fig. \ref{fig:init}.  For the sake of generality, we will consider two canonical configurations : the first includes an out-of-plane component of the magnetic field, which, for simplifying purpose, is assumed to be uniform and equal to $B_{gf}=1$ and the second one is a coplanar current layer ($B_{gf}=0$). Equations (\ref{eq:densityinit}-\ref{eq:magneticinit}) are defined in the hybrid coordinate system, replacing $y$ by $z$ defines them in the plane of the fully kinetic model. The domain is $64\delta_i$ in the downstream direction and $25.6\delta_i$ in the upstream direction, $y_0$ denotes half of the upstream size.  Because our system is periodic in the $x$ direction, the tearing instability developing within the current layer would eventually saturate with a fixed number of magnetic islands instead of ejecting them away and form a dominant X line where plasma jets are formed. To prevent this artificial evolution, and because we are mostly interested in the non-linear quasi-steady reconnection regime, we initially trigger magnetic reconnection with a local magnetic perturbation of amplitude $\delta B$, centered in the middle of the domain. To isolate the effect of the initial perturbation on the system from the one originating from lack of electron kinetic physics, the hybrid calculations are repeated with a perturbation twice as large as the one used in the original runs. The simulations presented in this paper are summarized in the table \ref{table:simulations}.


\begin{table}[htdp]
\caption{Summary of the simulations presented in this paper}
\begin{center}
\begin{tabular}{|c|c|c|c|}
\hline
Run & Model       & $B_{gf}$  & $\delta B$\\
        \hline
$FG$   & full PIC      & $1.0$  & 0.1\\
$FC$   & full PIC      & $0.0$  & 0.1\\
$HG_1$  & Hybrid PIC & $1.0$  & 0.1\\
$HC_1$  & Hybrid PIC & $0.0$  & 0.1\\
$HG_2$  & Hybrid PIC & $1.0$  & 0.2\\
$HC_2$ & Hybrid PIC & $0.0$  & 0.2\\
\hline
\end{tabular}
\end{center}
\label{table:simulations}
\end{table}%

The particles are initially loaded in Maxwellian distribution functions which locally have the density and temperature described above. Although it satisfies the pressure balance condition (\ref{eq:1dpressurebalance}), this initial condition is not a solution of the steady Vlasov-Maxwell system. As a result finite Larmor radius effect will modify the internal structure of the current layer \citep{Pritchett:2008ef,2011PhPl...18l2901A} where both electrons and ions find a new and self-consistent force balance.

 \begin{figure}
 \includegraphics[width=\linewidth]{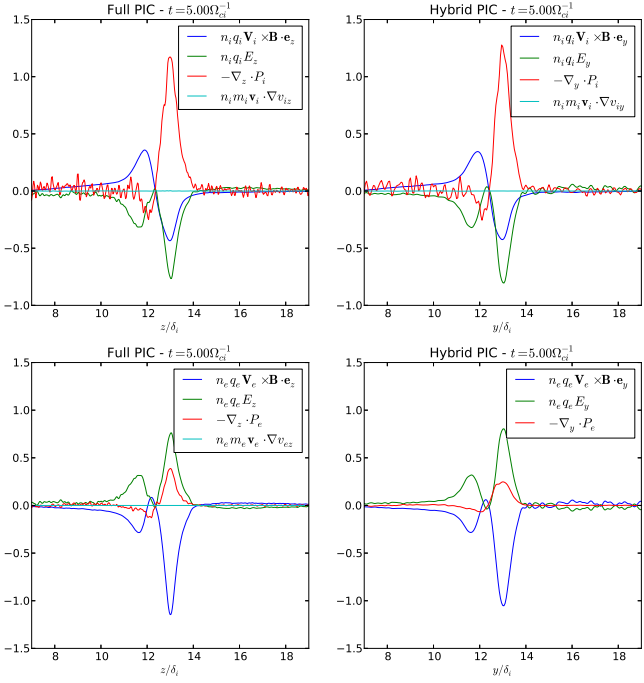}%
 \caption{\label{fig:init1} Force balance across the current sheet for the ions (top panels) and the electrons (bottom panels) at $t=5$ in the fully kinetic (left panels) and hybrid kinetic (right panels) simulations initialized with the coplanar configuration.}%
 \end{figure}

 \begin{figure}
 \includegraphics[width=\linewidth]{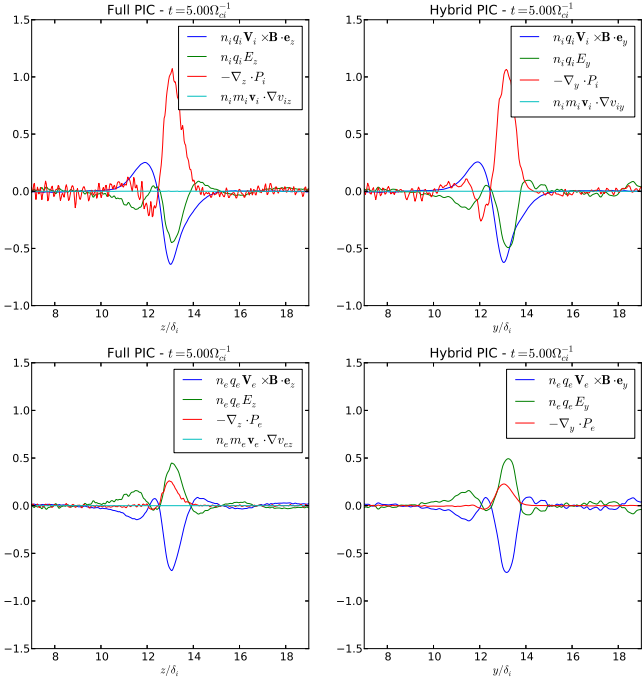}%
 \caption{\label{fig:init2} Force balance across the current sheet for the ions (top panels) and the electrons (bottom panels) at $t=5$ in the fully kinetic (left panels) and hybrid kinetic (right panels) simulations initialized with the guide field configuration.}%
 \end{figure}
 
From the modeling and fundamental viewpoint it is important to understand to what extent the kinetic nature of electrons participates to the establishment of this new equilibrium\citep{Alpers:1969dd,1996SSRv...76..251R,2003PhPl...10.2501M}, and what are the consequences of ignoring this part of the physics\citep{2012PhPl...19b2108B}. A good test is to compare the early evolution of the system between fully kinetic and hybrid kinetic simulations. Figures \ref{fig:init1} and \ref{fig:init2} show the ion and electron force balance at $t=5$ in both codes, in a cut in the cross current direction, for the coplanar and guide field configurations, respectively. The curves are obtained after an average of the force profiles between $x=0$ and $x=15$ to reduce the noise. As one can notice, the amplitude and variations of the forces are very similar in the fully kinetic and hybrid kinetic systems, for ions as well as for electrons, and for the coplanar and guide field cases. In the coplanar case, the ion equilibrium is mostly the result of a balance between the pressure force $-\boldsymbol{\nabla}\mathbf{P}_i$ and the electric force $en\mathbf{E}$, while in the guide field case the magnetic force $n_pq_p\mathbf{v}_i\times\mathbf{B}$ becomes as important as the electric force in balancing the pressure gradient force, indicating that ions are more magnetized. The electron equilibrium however, mostly concerns the balance between the electric force $n_eq_e\mathbf{E}$ and the magnetic force $n_eq_e\mathbf{v}_e\times\mathbf{B}$, and that in both the coplanar and guide field case. In a movie, available online, one can notice that the time scale over which the layer oscillates and emits waves are also identical in the hybrid and fully kinetic runs. As a result of this remarkable ressemblance, one can conclude that the self-consistent fluid equilibrium found by the system is mostly a consequence of the kinetic behavior of ions and not of the electrons, for which one can safely neglect the kinetic nature. Systems having an initially non-uniform temperature might show more differences and this have to be checked in future models. Let us notice, however, that the current sheet has initially a sub ion scale thickness, which is quite small compared to the Earth magnetopause current sheet for instance, and that the mass ratio used in this study is unrealistically large. As a consequence of these two effects combined, our model largely emphasizes the role of electron finite Larmor radius effects, if any. In a realistic configuration, considering the electrons as a fluid should therefore be an even better approximation.\\

\section{Reconnection rate}
\begin{figure*}
 \includegraphics[width=\linewidth]{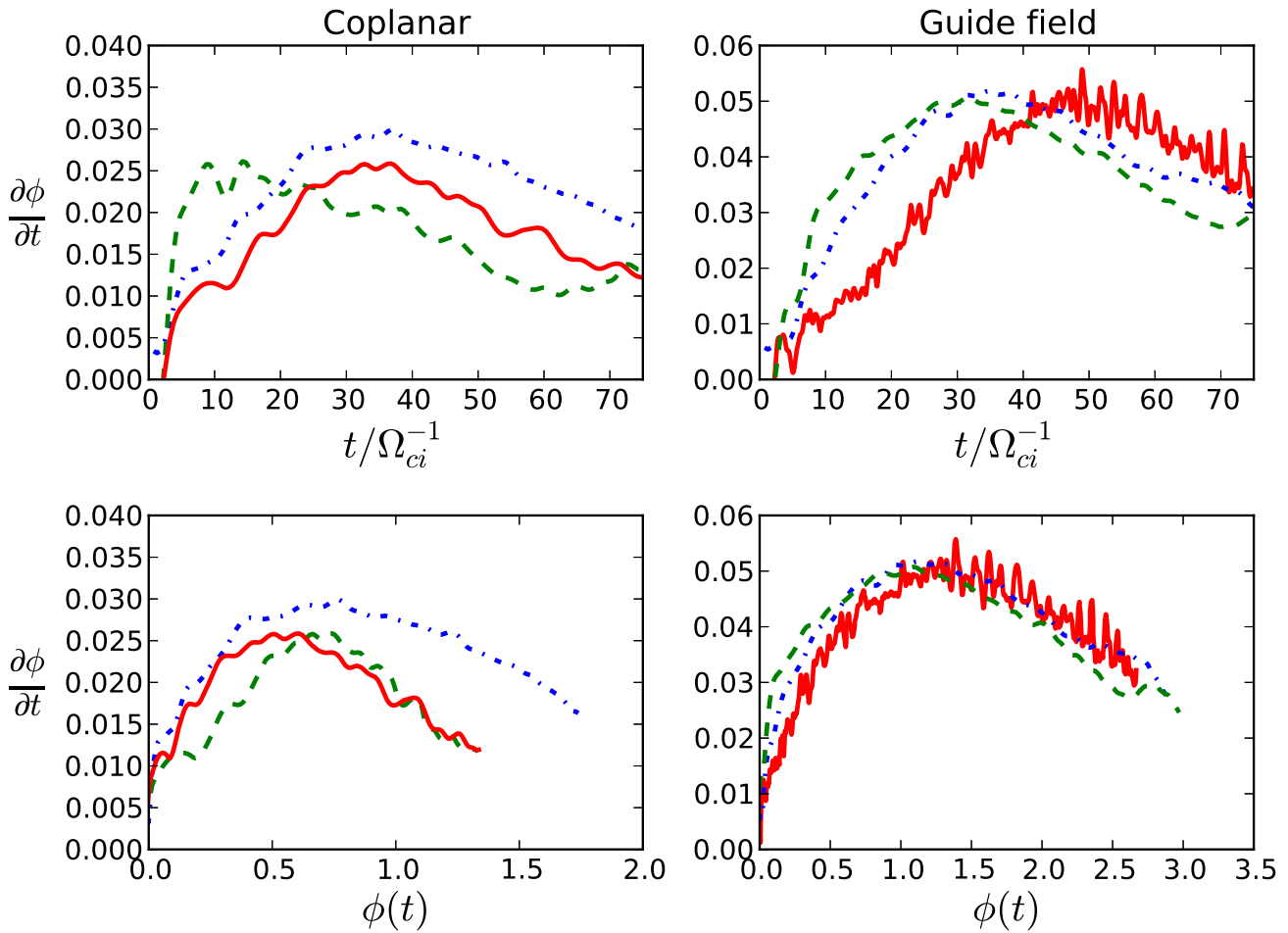}%
 \caption{\label{fig:rates} \textbf{Top panels} : Reconnection rate as a function of time for the coplanar (left) and guide field configurations (right). \textbf{Bottom panels} : Reconnection rate as a function of the reconnected flux $\phi$ for the coplanar (left) and guide field configurations (right). \textbf{On all panels}, the blue (dash-dot) and red (solid) curves are respectively obtained from the fully kinetic and hybrid runs sharing the same perturbation amplitude. The green curve (dash) is obtained from a hybrid run having a perturbation with an amplitude twice larger.}%
 \end{figure*}

 In this section, we compare the reconnection rate obtained from the hybrid and fully kinetic models.  The reconnection rate is the amount of magnetic flux reconnected per time unit. In a two-dimensional configuration, this definition corresponds to the electric field component perpendicular to the reconnection plane, measured at the X line. Technically, this electric field can be either directly measured in the simulation or be calculated as the time derivative of the out-of-plane component of the vector potential $\phi$ at the X line, which in 2D represents the variation of the reconnected flux. The latter method is preferred since it involves much less noise in the results than the former. Figure \ref{fig:rates} shows the reconnection rate for both numerical models in the coplanar and guide field configurations. The X line itself is found as being the saddle point of the out-of-plane component of the vector potential and the measure is averaged over an area of $0.04\delta_i^2$. The top panels show the time evolution of the reconnection electric field. The bottom panels show this rate as a function of the reconnected flux. All plots share the same tendency, where the rate first increases over a certain time interval, reaches a maximum value and slowly begins to decrease. This late phase is the consequence of the finite size of the domain, which limits the region over which the reconnected flux can be expelled and the amount of magnetic flux that can be reconnected.

Let us first look at the guide field configuration. On the top right panel, the hybrid and full PIC simulations sharing the same perturbation have a substantially different time evolution of the reconnection rate. However, it is worth noticing that their maximum reconnection rate is quite similar, the main difference being in the time period required to reach it. For the same perturbation amplitude, the hybrid model requires more time to reach the maximum rate than the fully kinetic model. Doubling the perturbation amplitude makes both models almost in phase. Because the two models differ only by the handling of the electron physics, these results suggest that the electron kinetic physics in one case, and the hyper-resistive dissipation in the other case, do not respond the same way to the initial perturbation. This comes from the fact that the early evolution of the process consists in a local collapse of the magnetic reversal down to the dissipation scale where convection and dissipation are in competition. Because the dissipation mechanism is different in both models, the time required to build the current sheet is different. In a fully kinetic simulation, one might expect this phase to depend on the electron to ion mass ratio. Fully kinetic codes are also subject to a larger noise than hybrid codes, which can also accelerate the development phase of the process. Looking at the top panel for the same configuration reveals that the reconnection rate as a function of the reconnected flux is identical for all simulations. Consequently, the role of the perturbation is just a time lag on the reconnection rate, which, fundamentally, rather depends on the phase of the process, i.e. on the upstream properties of the field and plasma, regardless of the kinetic nature of the electrons. 

The situation is a bit different for the simulation of the coplanar configuration. Here again the rate of the fully kinetic model increases faster than the one of the hybrid model, however the time lag of the latter is not as much as in the guide field case. As a result, doubling the perturbation amplitude, which, again, shortens the development phase, makes the hybrid model to reach the maximum before the fully kinetic one. As for the guide field scenario, the electron/dissipation physics does not respond the same way to the perturbation. However, contrary to the guide field case, it is unlikely that one finds a perturbation amplitude for which both models would be in phase with the same amplitude. Indeed, while they do not differ drastically, the maximum rate of the fully kinetic model is significantly larger than the hybrid one, which, itself, appears moreover not to depend on the amplitude of the perturbation. When looked as a function of the reconnected flux, the rate of both hybrid simulations appears to be more similar and separated from the one of the fully kinetic models. These observations suggest that in the coplanar case, the reconnection rate is sensitive to the nature of the electron physics, and that kinetic electrons apparently speeds up the process.

Finally, it is worth noticing the difference between the coplanar and guide field configurations. One can clearly see that the guide field runs have much larger reconnection rates than the coplanar ones. While guide field reconnection is as fast or slower than antiparallel reconnection in symmetric systems because in the latter the plasma is more compressible, our results suggest that this well accepted scenario cannot be extrapolated to asymmetric systems for which a better explanation has to be found. Because asymmetric current sheets are much more prevalent than symmetric ones, this result has broad consequences, in particular if one considers magnetopause reconnection. This finding will be studied in detail in a forthcoming paper. Let us remark at this point that the effect is seen in both hybrid and fully kinetic models, indicating that ions are the main contributor to this effect and not electrons.


 \begin{figure*}
 \includegraphics[width=\linewidth]{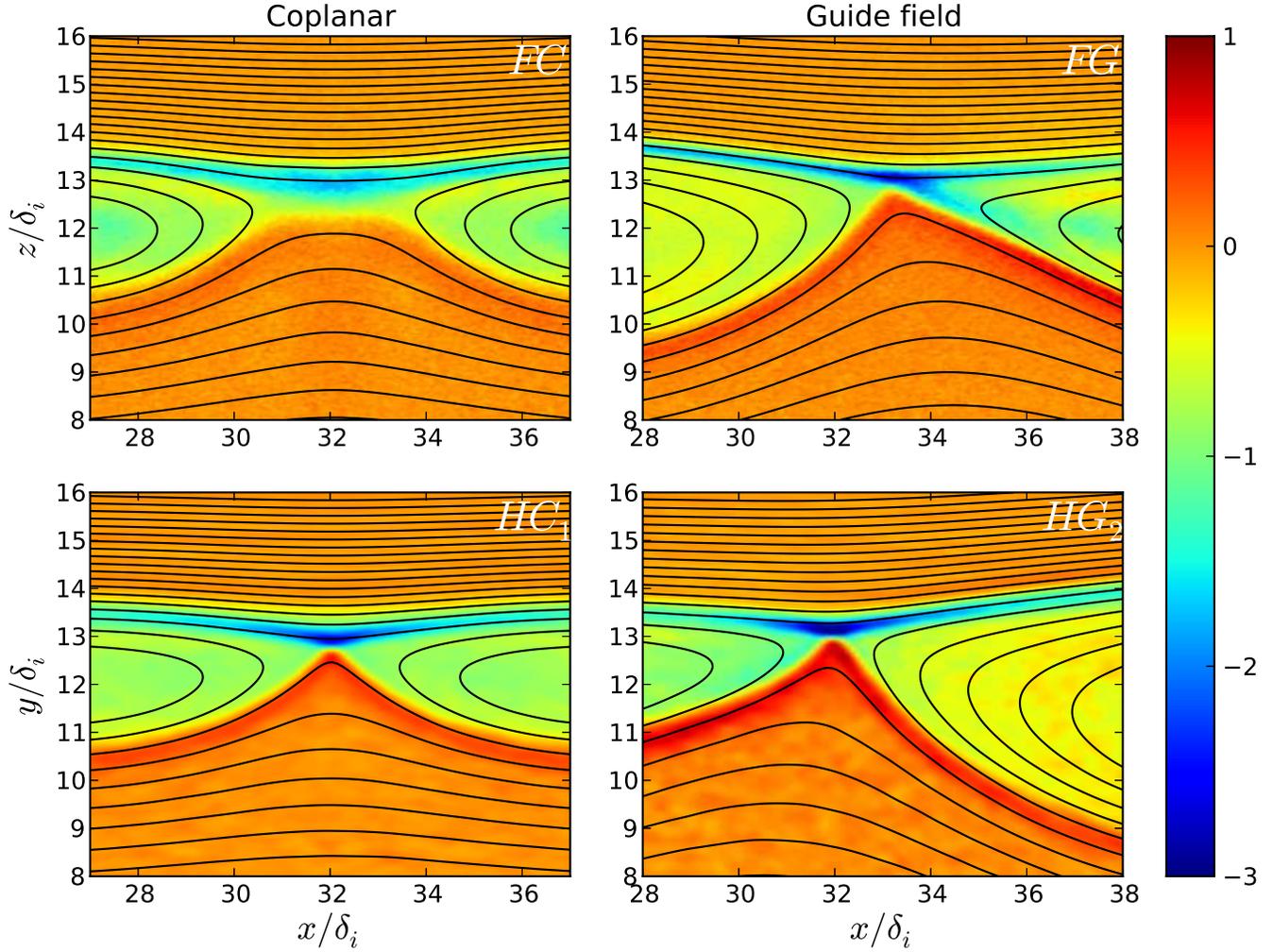}%
 \caption{\label{fig:jcolor} Out-of-plane current density at $t=35$ for the fully kinetic (top panels) and hybrid runs ($HG_2$ and $HC_1$, bottom panels) in the coplanar (left panels) and guide field (right panels) configurations. The current density of the fully kinetic model as been multiplied by $-1$ so that it can be compared more easily with the hybrid results obtained in the different coordinate system.}%
 \end{figure*}

 \begin{figure}
 \includegraphics[width=\linewidth]{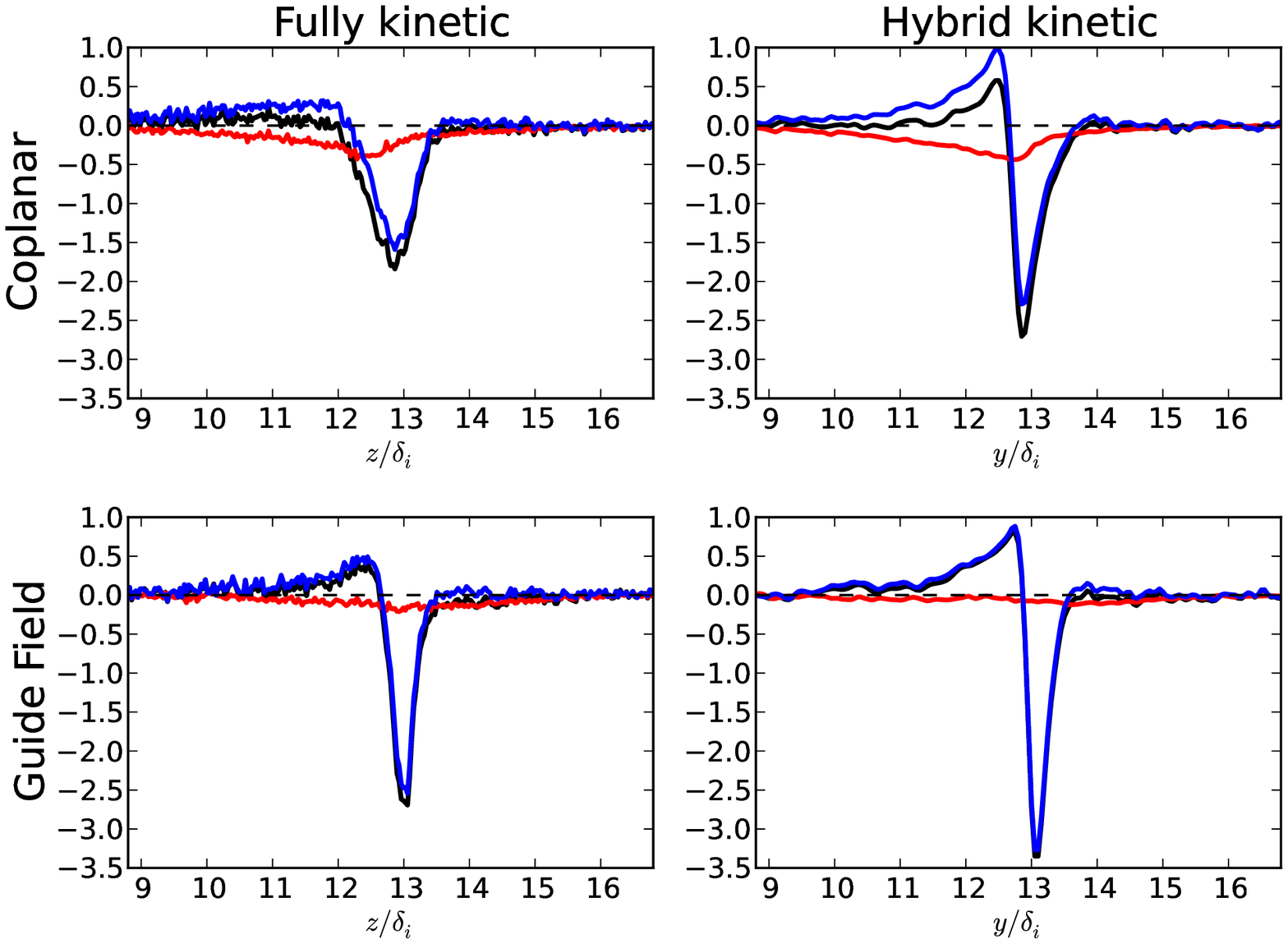}%
 \caption{\label{fig:jcut} Out-of-plane current density through the X point at $t=35$ for the fully kinetic and hybrid runs ($HG_2$ and $HC_1$) in the coplanar and guide field configurations. For all panels, the total current density is represented in black, the ion current density in red and the electron one in blue. As in Fig \ref{fig:jcolor} the current densities in the results obtained from the fully kinetic code have been multiplied by $-1$ to ease the comparison.}%
 \end{figure}

\section{Structure of the current sheet}
During the reconnection process, the magnetic field reversal at the X line itself is associated with a current sheet mainly supported by the lightest species in the plasma, namely the electrons. At this location, the current density is sustained through the acceleration of the electrons by the reconnection electric field and dissipated as the hot and fast electrons leave the layer by recoupling to reconnected field lines. A steady state is reached when these two effects balance each other\citep{2011SSRv..tmp...10H}. From the fluid viewpoint, this region of space is characterized by the competition between advection and dissipation mechanisms, and it is not clear to what extent the underlying kinetic behavior of the electrons controls the structure of the current sheet. Figure \ref{fig:jcolor} shows the out-of-plane current density for the hybrid and fully kinetic simulations in the coplanar and guide field configurations at $t=35$. Figure \ref{fig:jcut} represents a slice of the current density, the ion current density and the electron current density at the same time for the same simulations, along the upstream direction and through the X line. Both figures show the same features overall: the current sheet has a sub-ion scale mainly supported by an electron current, as expected. Looking at Fig. \ref{fig:jcolor} and in particular at the right panels, we can see that despite the difference in the symmetry originating from the different coordinate systems and the same guide field sign, both simulations look very similar from the ion scale down to the current sheet scale. Both models show a left right asymmetry that is a consequence of the initial guide field together with the Hall effect. The negative current density is strongly enhanced at the reconnection site and spreads on the top separatrices with more pronounced values on the big island side of the X line. Both models also show a current layer with an opposed sign on the bottom separatrix with values more pronounced on the smaller island side. The amplitude of the current density at the X line itself is roughly similar, but this will probably depend on the mass ratio in the fully kinetic model and on the value of the hyperresistivity in the hybrid kinetic model. The bottom panels of Fig. \ref{fig:jcut} reveal the very similar structure of the total current density between the two models. In both cases, the current sheet does not consists of a single peak but has a smaller peak of opposite sign on the weak field side. In both cases, the respective part of the current density supported by the ions and the electrons looks similar in amplitude and the spatial extent of the non-zero current density is also identical. One can notice that gradients look slightly sharper in the hybrid results than in the fully kinetic ones, this effect is probably controlled by the electron to ion mass ratio and the value of the hyperresistivity. \\

 \begin{figure*}
 \includegraphics[width=\linewidth]{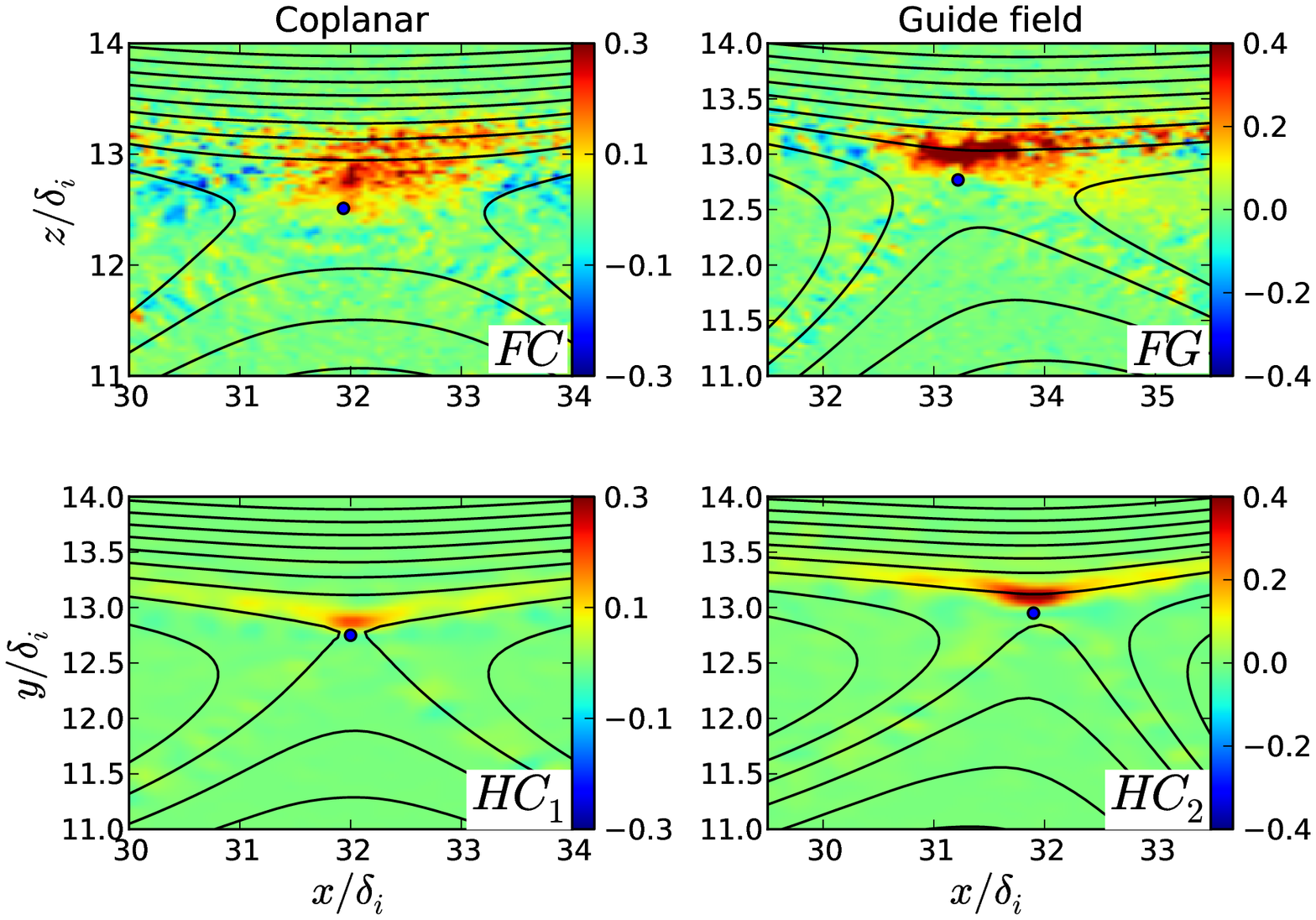}%
 \caption{\label{fig:dissipcolor} Dissipation measure\citep{2011PhRvL.106s5003Z} $D_e$ calculated at $t=35$ for the fully kinetic (top panels) and hybrid (bottom panels) in the coplanar(left panels) and guide field (right panels) configurations. Notice the color range is adjusted for each panel. The small blue circle denotes the position of the X point, localized as the saddle point of the magnetic flux function.}%
 \end{figure*}

As for the reconnection rate, the coplanar runs are again revealing more differences between the hybrid and fully kinetic models. The left panels of Fig. \ref{fig:jcolor} show that  the current sheet in the hybrid run is much more localized than its fully kinetic counterpart. In the fully kinetic simulation, the current sheet is both broader and longer in the upstream and downstream directions, respectively. As a result, the current density is much weaker than in the hybrid run. Looking at the top panels of Fig. \ref{fig:jcut}, one can see that the ion current density is very similar between both runs. It has the same amplitude and its variations are identical. The total current density is however different. While the hybrid one has roughly the same structure as in the guide field configuration, with the peak of opposite sign on the weak field side, the fully kinetic current density looks more symmetric, even if one can notice a slight increase on the weak field side too.\\

The fact that the structure of the hybrid current sheet differs from the fully kinetic one mainly in the coplanar case is a consequence of the lack of electron confinement in the vicinity of the X line. Like the ions, the electrons see a vastly different field on both sides of the current sheet, and they are less confined on the weak field side than on the strong field side. The guide field amplitude is large enough to change the ion magnetization in the current sheet, therefore it is also sufficient to magnetize the electrons, which are much lighter. As a result, the guide field scenario is less sensitive than the coplanar one to electron kinetic effects resulting from non-local mixing of populations. In the hybrid coplanar simulation however, electrons physics is local, which therefore leads to a more confined current sheet. Due to the unrealistic electron mass used in the simulations, we expect that lighter electrons will result in their better confinement and less differences between the hybrid and fully kinetic models.

\section{Motion of the X line and dissipation region}

Our hybrid model does not include non-gyrotropic, or even anisotropic electron pressure. It is therefore interesting to tests to what extent, the features for which these effects have been identified as a key ingredients, differ between the hybrid and fully kinetic runs. We choose to investigate two of those features. The first one is the extent of the dissipation region around the X line. It has recently been shown\citep{2008PhPl...15k2102H} that the electron scale current layer was not necessarily a good proxy of a dissipative process. Another recent study\citep{2011PhRvL.106s5003Z} then proposed a scalar quantity $D_e$ to measure non-ideal energy transfers from the electromagnetic fields to the plasma. In a collisionless environment, irreversible energy transfers are the result of complicated mixing is phase space that macroscopically appear in the non-gyrotropic components of the electron pressure tensor. In the hybrid model, these terms are missing, however, they are modeled by a simple uniform hyperresistivity. Fig. \ref{fig:dissipcolor} shows the dissipation measure $D_e$ in the hybrid and fully kinetic runs, for the coplanar and guide field configurations. $D_e$ is strongly localized in the reconnection region in both hybrid and fully kinetic runs, and is better confined in the hybrid runs, since no electron finite Larmor radius effect is occurring. We can again notice the confinement effect the guide field has on electrons, as the dissipation region is, in this configuration, better localized than in the coplanar case. One puzzling result is that the strongest dissipation seems to be shifted from the actual position of the X line, it appears to be collocated with the maximum of the current density (Fig. \ref{fig:jcolor}). This could have already been notice in a previous work\citep{2011PhRvL.106s5003Z} were the same initial condition was used. These features are seen in fully kinetic and hybrid runs, which indicates that they are not simply resulting from an electron kinetic effect. A detailed investigation of the structure of the dissipation region will be the topic of a future study.

\begin{figure}
 \includegraphics[width=\linewidth]{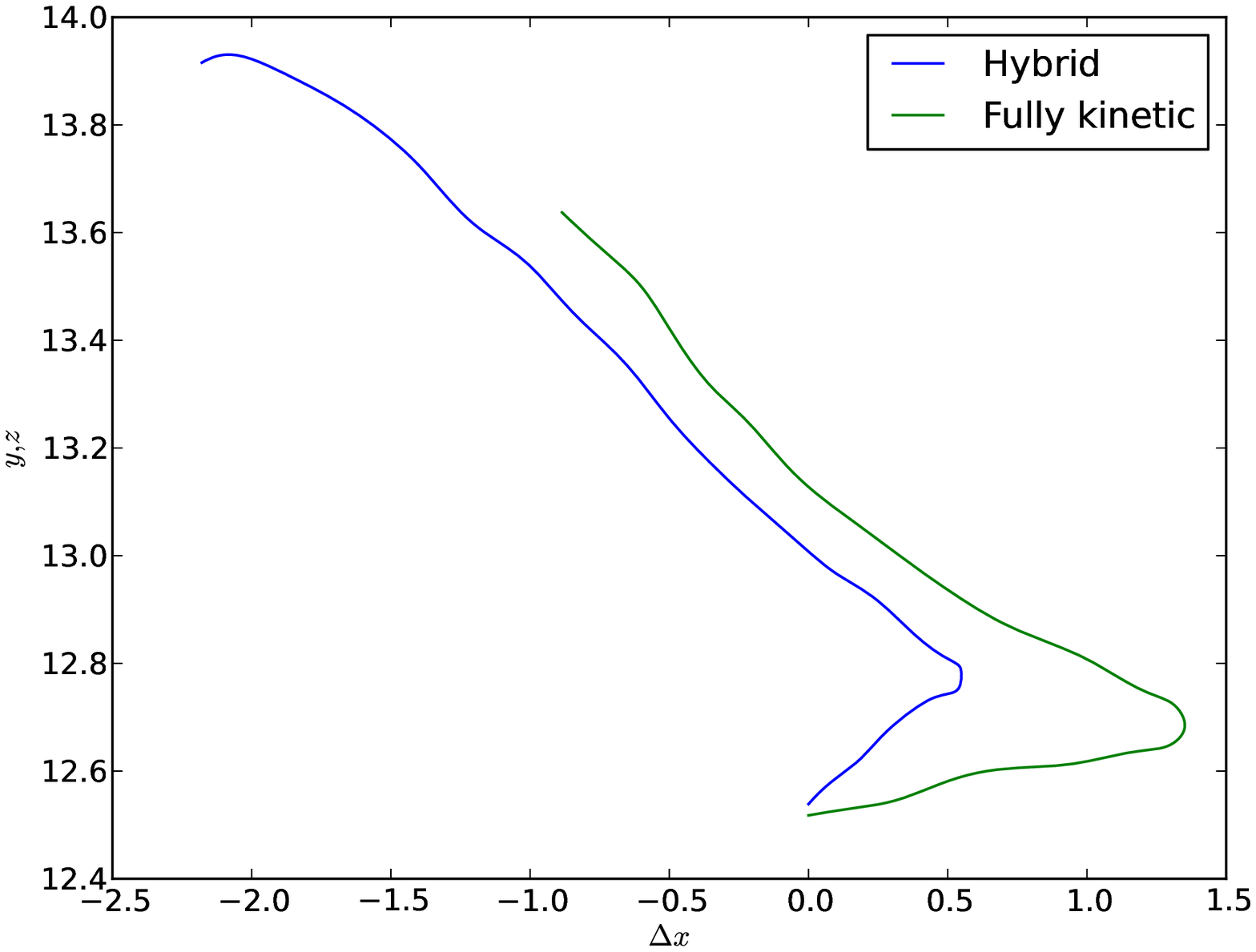}%
 \caption{\label{fig:xlinemotion}Position $\Delta x$ of the X line in the reconnection plane for the hybrid and fully kinetic model with respect to its initial position. To ease the comparison, the blue curve shows the mirror with respect to $x=0$ from the actual position obtained from the hybrid model, which, otherwise be the opposite because of the different coordinate system used.}%
 \end{figure}

The second feature we will discuss is the motion of the X line often seen in asymmetric configuration with a guide field and often associated to an electron diamagnetic effect\citep{2003JGRA..108.1218S}, i.e. an electron pressure effet. As one can see in the figures \ref{fig:jcolor} and \ref{fig:dissipcolor}, the X line in the fully kinetic model is approximately located at $x=33.2$, whereas the hybrid X line appears closer to its original location ($x=32$). To investigate whether the two models have a qualitative difference regarding the motion of the X line, we show, on Fig.\ref{fig:xlinemotion}, the position, in the reconnection plane, of the X line, in the hybrid and fully kinetic models. Noticing that the apparent symmetry is a result of the different coordinate system, one can see a very strong similarity in the motion of the X line. First, the X line does not move very far from its initial location. Then, and more surprisingly, it begins to move in one direction and then starts moving in the other one until the end of the calculation. The effect is quantitatively seen in both hybrid and fully kinetic models. Small quantitative differences can be seen, the most important being that the fully kinetic X line travels farther than the hybrid one before changing direction a little bit later. There is, consequently, a slight timing difference for the reversal time ($t\approx 27$ in the fully kinetic run and $t\approx 20$ in the hybrid one), which results, for the hybrid run, in an X line being coincidentally close to its original position at the time ($t=35$) when Fig. \ref{fig:jcolor} and \ref{fig:dissipcolor} are made. In both models, the X line also moves upward, with a slightly faster speed in the hybrid model. It is not clear why the X line changes its direction. The electron and ion diamagnetic drift speeds (not shown) are in opposite directions all along the simulation. However, their values around the X line exceed by far the velocity of the X line motion. It is possible that the first phase corresponds to an influence of the initial perturbation while the second phase is more a self-consistent feature of the reconnection process. A more detailled investigation is beyond the topic of the present paper, which aims to focus on the overall evolution of asymmetric reconnection.

\section{Summary and discussion}
In this paper we have compared hybrid and fully kinetic simulations of asymmetric magnetic reconnection to investigate the role of the kinetic nature of the electrons in the overall evolution of the system. We have chosen two configurations, one with a coplanar current sheet and a second with a uniform guide field. The initial condition is a fluid pressure balance with locally Maxwellian distribution functions. This initial state being not a Vlasov equilibrium, finite Larmor radius effects rapidly change the internal structure of the current sheet and establish a new self consistent force balance while waves propagate away. We have shown that the force balance found by the system is identical in the full PIC and hybrid models, in amplitude, spatial and temporal scales, indicating that the process is controlled by the kinetic behavior of the ions and not by kinetic electrons. This should also be inspected and confirmed in other systems, having an initial temperature gradient for instance, however the very small thickness of our initial current sheet and the large electron mass used in this system indicate that the kinetic nature of the electrons might not be critical in real systems like the quiet magnetopause where gradients are usually at the ion scale.\\

We have also shown that the reconnection rate obtained from the hybrid and fully kinetic results are fairly similar. One important difference is the response of the two models to the initial perturbation, the hybrid model taking more time to reach the maximum reconnection rate than the fully kinetic model. This effect originates from the distinct dissipation mechanism, which plays a critical role in the establishment of the current sheet at the reconnection site and also possibly of the electrostatic noise of fully kinetic codes larger than in hybrid codes. Overall, the initial perturbation has been shown to affect the time dependance of the reconnection rate but its dependance on the phase of the process, i.e. on the upcoming magnetic field and plasma properties, stays unchanged. When plotted as a function of the reconnected flux, the difference in the reconnection rates between hybrid and fully kinetic models for the guide field configuration is negligible. However, the coplanar runs show some differences: the rate in the fully kinetic model being somehow larger than the hybrid one. These similarities and differences of the hybrid and fully kinetic models for the guide field and coplanar configurations, respectively, have also been shown in the structure of the out-of-plane current density. If it is very similar for the guide field case, it is appreciably different in the coplanar case. The guide field configuration leads to more confinement of both species inside the current layer, in particular of electrons. However this confinement is a kinetic process due to the mixing and the bouncing of particles inside the magnetic reversal, it can therefore occur for ions in both models but for electrons only in the fully kinetic model. As a result, without a guide field the ion current density gets broader in both models but the electron current density stays localized in the hybrid one, which changes the total current sheet structure. We expect the difference between the two models to diminish as the electron mass gets smaller, which should be tested in future studies. A last important point is the difference of the reconnection rate between the guide field system and the coplanar one. The guide field case is considerably faster than  the coplanar system, which is surprisingly opposed to the common understanding of the effect of a guide field on reconnection. The fact that the hybrid and fully kinetic models both show the same tendency, to the minor differences explained above, indicates that ions are the primary responsible for this effect. This finding has broad consequences, for instance in the debate of whether magnetopause reconnection would prefer guide field or coplanar configurations, and will be analyzed in detail in a forthcoming paper. \\

As a last step, we have studied the extent of dissipative energy transfers and the motion of the X line, both effects being, associated to the electron pressure tensor. We have shown that the dissipation measure proposed recently\citep{2011PhRvL.106s5003Z} is well localized in the reconnection region in both hybrid and fully kinetic models. In all models, the structure is slightly shifted from the actual X line and appears collocated with the maximum of the current density. In the presence of a guide field, the X line has been observed to move slowly although its motion has not been found related to an ion or electron diamagnetic effect. These results emphasize the need for a detailed investigation of the different processes occurring in the vicinity of the X line, which appear to differ substantially from the present understanding inferred from symmetric models. Future studies should also consider larger systems and longer simulations times as they might possibly reveal unexpected behaviors as it did for reconnection in symmetric systems.













%




%











%







\begin{acknowledgments}

Three of us (N.A., C.B. and R.E.) acknowledge support from the NASA postdoctoral program. M.H. acknowledge support from the theory and modeling group of NASA's MMS.
\end{acknowledgments}


%

\end{document}